\documentclass[10pt]{iopart}
\usepackage[T1]{fontenc}
\usepackage{epsfig}
\usepackage{graphicx}
\usepackage{dcolumn} 
\usepackage{tabularx} 
\usepackage{array}       
\usepackage{tabularx}
\usepackage{dcolumn} 
\usepackage{multirow}
\usepackage{siunitx}
\usepackage{booktabs}
\newcolumntype{Y}{>{\centering\arraybackslash}X}
\usepackage{upgreek} 
\usepackage{bm}
\usepackage{iopams}

\usepackage[usenames,dvipsnames]{xcolor}

\newcommand{\ket}[1]{\ensuremath{\left|#1\right\rangle}}

\begin{document}

\title[]{Efficient cooling of high-angular-momentum atoms}

\author{Logan E. Hillberry$^1$, Dmitry Budker$^{2,3}$, Simon M. Rochester$^4$, Mark G. Raizen$^1$}

\address{$^1$ Department of Physics, The University of Texas at Austin, Austin, Texas, 78712, USA}
\address{$^2$Johannes Gutenberg-Universit{\"a}t Mainz, Helmholtz-Institut Mainz, GSI Helmholtzzentrum f{\"u}r Schwerionenforschung, 55128 Mainz, Germany}
\address{$^3$ Department of Physics, University of California, Berkeley, California 94720, USA}
\address{$^4$Rochester Scientific, LLC, El Cerrito, California 94530, USA}
\ead{lhillber@utexas.edu}
\vspace{10pt}
\begin{indented}
\item[]May 2023
\end{indented}

\begin{abstract}
We propose a highly efficient and fast method of translational cooling for high-angular-momentum atoms. Optical pumping and stimulated transitions, combined with magnetic forces, can be used to compress phase-space density, and the efficiency of each compression step increases with the angular momentum. Entropy is removed by spontaneously emitted photons, and particle number is conserved. This method may be an attractive alternative to evaporative cooling of atoms and possibly molecules in order to produce quantum degenerate gases.
\end{abstract}

%
\submitto{\jpb}
%
\maketitle
%
\ioptwocol
\section{Introduction}

Laser cooling, first proposed almost half a century ago, remains the standard approach for producing ultracold atoms \cite{metcalf1999laser,schreck2021laser}.  This method relies on momentum transfer from light to atoms as photons are repeatedly scattered, enabling the production and study of ultracold atomic gases. Many improvements on basic laser cooling have advanced the state of the art, including sub-Doppler Sisyphus cooling \cite{Lett1988, Dalibard1989,Ungar1989,Weiss1989,steane1991laser}, and sub-recoil degenerate Raman sideband cooling cooling \cite{Vuletic1998,Kerman2000}.  

While laser cooling works extremely well, the requirement of a closed, two-level transition has limited the applicability of the method to a subset of elements in the periodic table.  For those atoms, after many years of refinement, laser cooling has reached saturation in its performance due to multiple scattering of resonant photons which create an effective repulsive interaction between the atoms, pushing them apart.  An important figure-of-merit is the phase-space density, a dimensionless parameter which is the product of number density and the third power of the average de Broglie wavelength.  Laser cooling typically produces a phase-space density of 10$^{-6}$.  This is also the starting point for creating quantum degenerate gasses with order-unity phase-space density through evaporative cooling in a trap \cite{hess1986evaporative,masuhara1988evaporative}, and the creation of the so-called atom laser \cite{Mewes1997,Andrews1997,Bloch1999}. Evaporative cooling is even more restrictive than laser cooling, as it relies on elastic collisions between atoms to maintain thermal equilibrium as the hottest atoms are ejected.  Inelastic channels create unwanted losses and often make evaporative cooling impossible.  Even when working optimally, evaporative cooling is a slow process and results in a significant loss of atom number. 

\begin{figure}[htbp]
\includegraphics[width=\columnwidth]{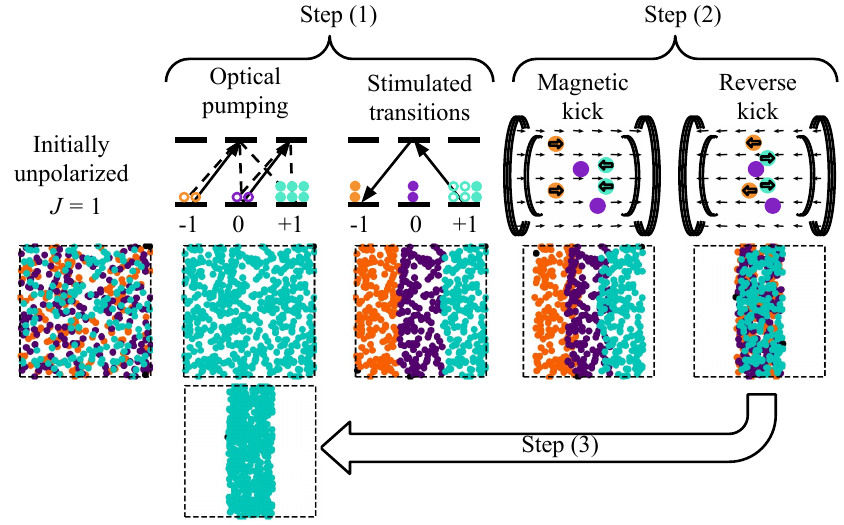}
\caption{A schematic depiction of the MOP-cooling sequence for ensembles with angular momentum $J=1$. The colored circles represent atoms, initially trapped in a flat, hard-wall potential. The colors represent the atom's magnetic state (orange: $m_J=-1$, purple: $m_J=0$,  teal: $m_J=+1$). A cycle begins with optically pumping all atoms to the same state. Then, a sequence of stimulated transitions correlate the magnetic states with position along the direction of compression. A sequence of one-dimensional magnetic kicks pushes atoms of oppositely-signed magnetic states together. The cycle is closed by optically pumping the compressed atoms back into the same magnetic state. A cycle's compression factor increases with the number of available magnetic states.} 
\label{fig:schematic}
\end{figure}

In recent years, alternative approaches to producing cold atoms and molecules were developed (see \cite{raizen2009} and references therein).  The starting point for much of this work is a supersonic molecular beam where desired atoms or molecules can be entrained in the flow and stopped in a series of pulsed magnetic or electric fields.  Alternatively, cold atoms and molecules are produced by buffer-gas cooling \cite{egorov2002buffer}.  After stopping, these atoms and molecules can be trapped, typically in a magnetic field configuration that confines the low-field seekers to the center of the trap.  In parallel, cooling of atoms with a one-way wall was proposed and demonstrated \cite{Raizen2005,Price2008,thorn2008experimental}, and relies on photon entropy, not momentum as in laser cooling.  The one-way wall is the first practical realization of Maxwell’s demon for an ensemble of atoms.  While one-way-wall cooling demonstrated a large increase in phase-space density, it did not conserve atom number. To address this limitation, we proposed \cite{Raizen2014} a variation which we called magneto-optical (MOP) cooling, relying on cycles of state preparation and magnetic kicks.  

In this paper, we present a generalized and highly efficient version of MOP cooling that can work for high-angular-momentum atoms and offers an attractive alternative to laser cooling and evaporative cooling. Quantum gases of atoms with high angular momentum have become of increasing scientific interest over the last 20 years \cite{chomaz2022dipolar} making our method timely and relevant to a growing community. As depicted schematically in Fig.\,\ref{fig:schematic}, and described in detail below, the larger internal-state space of high-angular-momentum atoms may be leveraged to yield a greater compression of an ensemble in real space. This is in contrast to our original proposal \cite{Raizen2014} that focused on spin-1/2 systems.
Our method will help bridge the existing gap of approximately six orders of magnitude between the phase-space density limits of standard sub-Doppler cooling and evaporative cooling, thereby promising larger ultracold samples.

In the following section, our simulation methodology is described, followed by a brief review of MOP cooling for a spin-1/2 system. Then MOP cooling is generalized for high-angular-momentum atoms.  We conclude the paper with a discussion of possible limitations to our new method, how those limitations may be overcome, and the significance of MOP cooling in the atomic physics toolbox. The subsequent appendix presents a one-dimensional model of MOP cooling as an alternative to the detailed simulations that follow.

\section{MOP cooling}
MOP cooling is a conceptually new method that does not rely on the momentum of the photon, making it completely different from laser cooling. The key benefit of this approach is its universality and simplicity, since it relies only on internal magnetic state preparation and magnetic forces from pulsed coils. MOP cooling may be interpreted as a technique that transforms the cooling of internal degrees of freedom via optical pumping \cite{kastler1950quelques} to external degrees of freedom. Through numeric simulations, we will evaluate the efficacy of MOP cooling for atoms with different numbers of internal states.

 Our MOP-cooling simulations track the three-dimensional positions ${\bf x}$, velocities ${\bf v}$, and magnetic states $m_J\in\{-2J, -2J+1,\ldots,2J\}$ of a sample of $N=10^5$ atoms, each with total angular-momentum quantum number $J$. The internal magnetic state of each atom is initialized in accordance with the MOP-cooling cycle (discussed further below).  Positions are initialized from a flat distribution of a 0.5\,cm  width. Velocities are initialized from the Maxwell-Boltzmann distribution corresponding to a temperature of $25 \times T_{\rm rec}$ where $T_{\rm rec}=h^2/M k_{\rm B}\lambda^2$ is the recoil temperature imposed by, e.g., a magneto-optical trap operating at wavelength $\lambda$ to trap a species of mass $M$. Here, $h$ is Planck's constant and $k_{\rm B}$ is Boltzmann's constant. 
 
 A fourth-order Runge-Kutta algorithm updates the position and velocity of each atom subject to the force ${\bf F} (t) = -m_J g_J \mu_{\rm B} \nabla \left | {\bf B}({\bf x}, t)\right|$ where $g_J$ is the Land{\'e} $g$-factor of the atom's ground state, $\mu_{\rm B}$ is the Bohr magneton, and ${\bf B}$ is the pulsed magnetic field arranged to provide the one-dimensional kick. The simulation time step is one microsecond. Table\,\ref{tab:atoms} reports the atomic properties used in this study.

\setlength{\extrarowheight}{.3em}
\begin{table}[htb]
    \centering
    \caption{Atomic properties used for simulation purposes. The Ref.~column provides an example of magneto-optical trap operation for each species with a corresponding recoil temperature $T_{\rm rec}$. Values for $g_J$ are provided in \cite{NIST_ASD}, except for Li for which we adopt the electron's value.}
    \begin{tabularx}{\columnwidth}{lS[table-format=2.2]cS[table-format=1.5]S[table-format=2.1]r}
         {\multirow{2}{*}{Atom}} &  {$M$} & {\multirow{2}{*}{$J$}} &{\multirow{2}{*}{$g_J$}}  & {$T_{\rm rec}$} & \multirow{2}{*}{{Ref.}} \\
           & {($10^{-26}$ kg)} & & & {$({\rm \upmu K})$} & \\ \hline
         {Li} & 1.15 & {1/2} & 2.00232  & 6.1 & \cite{hulet2020}\\
         {Cr} & 8.63 & {3}   & 2.00183 & 2.0 & \cite{bradley2000}\\ 
         {Er} & 27.83 & {6}   & 1.16381 & 0.3  & \cite{frisch2012} \\ 
         {Dy} & 26.98 & {8}   & 1.24159 & 0.3  & \cite{lunden2020} \\ 
    \end{tabularx}
    \label{tab:atoms}
\end{table}
 
 The magnetic field is modeled as two sets of coaxial coil pairs, one in the Maxwell configuration to provide a strong gradient \cite{turner_minimum_1998} and the other in the Helmholtz configuration to shift the zero-crossing of the field away from the ensemble's center, thereby providing a nearly-one-dimensional kick. We use superposition of the exact solution for a current-carrying loop to model the full coil geometry. ${\bf B}$ is evaluated on a dense grid for unit current. Vector interpolation allows us to evaluate $\nabla \left | {\bf B}({\bf x}) \right |$ at arbitrary positions.
Time-dependent current pulses are modeled by scaling the field gradient interpolation result by $I(t) = I_0 \sin[\pi(t-t_0) / \tau]$ for $t \in [t_0, t_0+\tau]$ and $I(t)=0$ otherwise, where $I_0$ is the peak current, $t_0$ is the pulse delay, and $\tau$ is the pulse width. In this paper, we set $\tau=300~{\rm \upmu s}$.

\begin{figure}[htb] 
\includegraphics[width=3.375in]{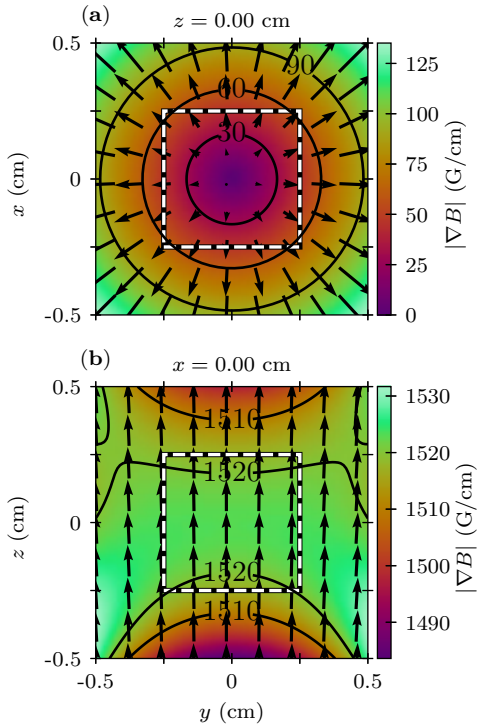}
\caption{Two-dimensional slices of the magnetic field gradient used for MOP cooling simulations, evaluated at peak current, and shown (a) transverse-to and (b) along the kicking direction. The black-and-white dashed line marks the initial extent of the atomic cloud. The magnitude of the total field $|{\bf B}|$ varies primarily along $z$ in a nearly-linear fashion.
}\label{fig:B_slices}
\end{figure}

The simulated coil parameters are as follows. There are 7 turns $\times$ 2 layers, or 14 loops per Helmholtz coil, each with a nominal radius of $R_{\rm HH} = 3.5\,\rm{cm}$, axially-separated by $R_{\rm HH}$, and carrying identically-oriented currents . There are 5 turns $\times$ 2 layers, or 10 loops per Maxwell coil, each with a nominal radius of $R_{\rm HH}/\sqrt{3}$, separated by $R_{\rm HH}$, and carrying oppositely-oriented currents. The peak currents are $I_{0,{\rm HH}}=1000$\,A for the Helmholtz coils and $I_{0,{\rm M}}=800$\,A for the Maxwell coils. The shared midpoint between the coil pairs is displaced by 0.2\,cm in the positive $z$-direction from the initial center of mass of the atomic sample (taken as the origin of the coordinate system). We find that this region provides a more uniform and one-dimensional kick. With these coil parameters, the peak field gradients are approximately 1520\,G/cm along the kick direction and 45\,G/cm in the transverse directions. Two-dimensional slices of the magnetic field gradient used in our simulations are shown in Figure \ref{fig:B_slices}.

\begin{figure*}[t] 
\includegraphics[width=6.75in]{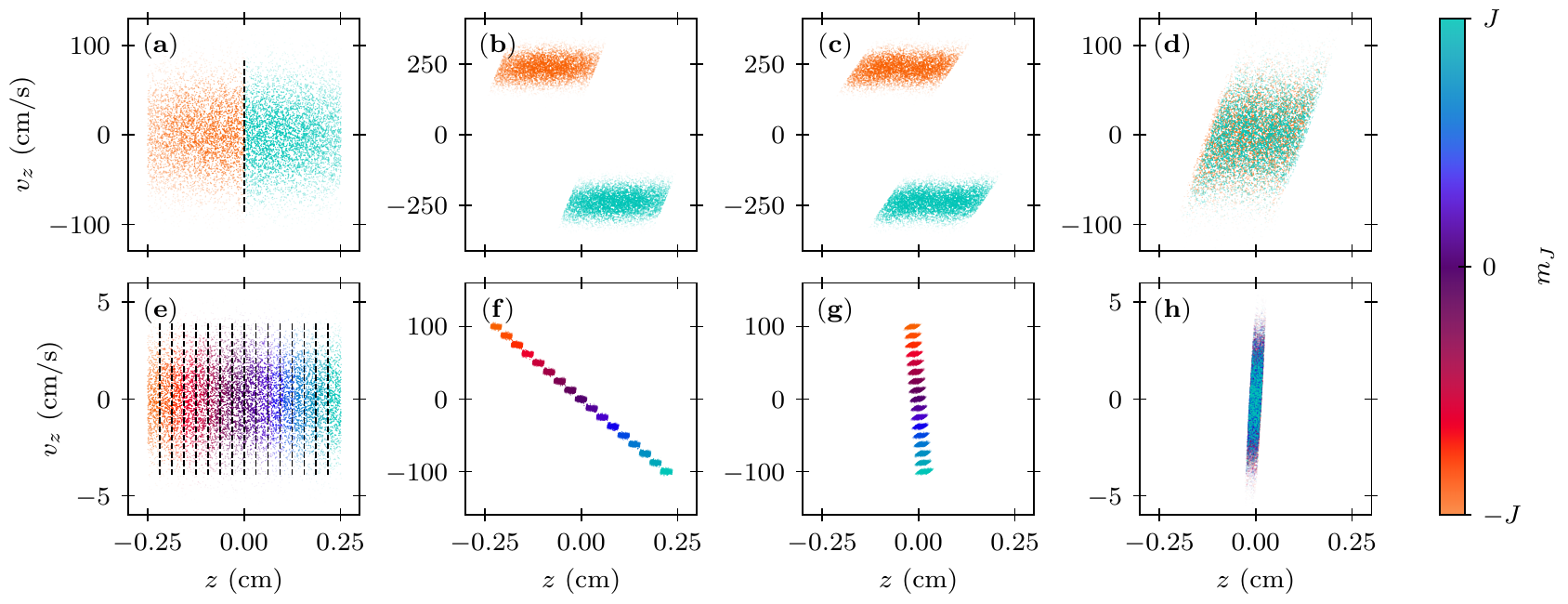}
\caption{
MOP cooling simulations for Li ({\bf a}-{\bf d}) and Dy ({\bf e}-{\bf h}) visualized in  phase-space. Each column of plots represents a snapshot during the magnetic kicks required for a cycle. Initially, the magnetic states are correlated with position along the $z$-axis through optical pumping followed by spatially-resolved coherent population transfer via stimulated transitions. The vertical dashed black lines in panels ({\bf a}) and ({\bf e}) mark the ideal boundary between different magnetic states. In panels ({\bf b}) and ({\bf f}), a one-dimensional magnetic kick accelerates atoms to a velocity that is proportional to their magnetic state. After waiting an optimized delay time the phase-space distribution has been compressed in real space but remains extended in velocity space, as shown in panels ({\bf c}) and ({\bf g}). In panels ({\bf d}) and ({\bf f}), we show a reverse kick returning the atom's velocity distribution to near its original extent while preserving the spatial compression. More precisely, we find the relative difference between standard deviations of the ensembles' initial and final velocity distribution is below 0.2\% for all coordinates and all four species simulated.
}\label{fig:phasespace}
\end{figure*}

For a specific example, consider atomic lithium (Li) trapped using standard techniques \cite{hulet2020}. At sufficient magnetic fields, the electronic spin decouples from the nuclear spin yielding a total angular momentum of $J=1/2$; the two electronic $m_J$ states are denoted \ket{1/2} and \ket{-1/2} and it is in this high-field regime that Li may be MOP-cooled as originally discussed in \cite{Raizen2014}. The cooling sequence starts with suddenly turning off the trap so that the atoms are free. In step (1) of MOP cooling, all of the atoms are optically pumped to the \ket{1/2} state. Then, half of the cloud is transferred to the \ket{-1/2} state by stimulated transitions, i.e., stimulated Raman adiabatic passage (STIRAP) sequences \cite{Bergmann1998,Bergmann_2019}, thereby creating two spatially-distinct populations [see Fig.\,\ref{fig:phasespace} ({\bf a})].  In step (2), a magnetic-field-gradient kick is applied to the cloud, thereby causing the two halves to merge [Fig.\,\ref{fig:phasespace} ({\bf b}-{\bf c})], and then a reverse kick returns the atoms to their original velocity distribution [Fig.\,\ref{fig:phasespace} ({\bf d})].  As demonstrated experimentally in \cite{Melin2019}, such magnetic kicks can be applied along a single axis while minimally affecting the other two dimensions.  In step (3), all atoms are optically-pumped back to the \ket{1/2} state, thereby completing the cooling cycle. In principle, a factor of $2\times$ in phase-space compression is possible. 

\begin{figure}[htbp] 
\includegraphics[width=\columnwidth]{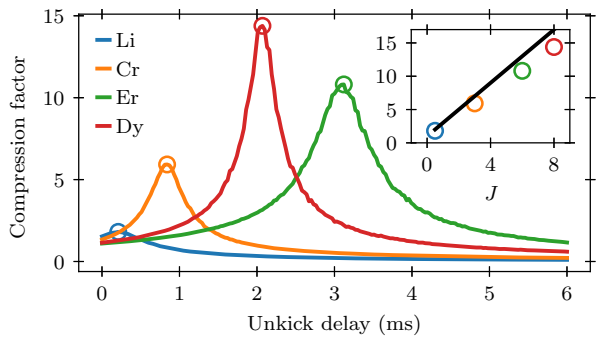}
\caption{Optimizing the wait time between the MOP cooling kick and unkick according to the compression factor. The open circle marks the optimal delay time for each species subject to the initial conditions and magnetic forces described in the text. The inset shows the peak compression factor vs the species' angular momentum $J$, compared to the geometric limit $2J+1$ (black line).
}\label{fig:optimize_wait}
\end{figure}
 
We now generalize the method to a system with an arbitrary total angular momentum $J$. There are $2J+1$ states in this case, and we assume that the atoms are trapped in a hard-walled flat potential. Just as in the case of Li above,  we turn off the trap to start the cooling sequence.  Using optical pumping and stimulated transitions in step (1), the cloud is divided into $2J+1$ equally-sized components, where the leftmost section is prepared in state $\ket{J,-J}$, then $\ket{J,-J+1}$, and so on, to the rightmost section in state $\ket{J,J}$.  In step (2), a magnetic field gradient kick is applied to the cloud, causing each sub-section to move at a velocity that is proportional to its distance from the center-plane.  Thus, the stretched states $\ket{J,\pm J}$ will move the fastest towards the center, and lower magnetic-quantum-numbered sections will move slower.  The cloud will collapse to a single section after an optimal delay time, and a reverse magnetic kick will restore the original velocity distributions. Finally, in step (3), all atoms are optically pumped to the $\ket{J,J}$ state, and the cloud can be re-trapped. Ideally, the cloud has been compressed by a factor of $2J+1$.

Figures \ref{fig:phasespace} ({\bf e} - {\bf h}) show snapshots of the simulated  phase-space for MOP cooling of Dy atoms ($J=8$) after state preparation (step 1) and during magnetic kicking (step 2). Comparing the final spatial distribution of Li [Fig.\,\ref{fig:phasespace} ({\bf d})] to that of Dy [Fig.\,\ref{fig:phasespace} ({\bf h})] clearly demonstrates how MOP cooling may leverage high-angular-momentum atoms for efficient phase-space compression. 

The delay time between kicks is optimized in our numeric simulations and the results are shown for a variety of atoms in Fig.\,\ref{fig:optimize_wait}. As a figure of merit, we compute the compression factor as the ratio of standard deviations between the initial and final $z$-coordinate distributions of the cloud. The inset of Fig.\,\ref{fig:optimize_wait} shows the peak compression factor for each species is bound by the geometric limit $2J+1$. In the following section we consider some limitations of MOP cooling and how they may be overcome in an experiment.

\section{Discussion}
In MOP cooling, a maximum compression factor of $2J+1$ per cycle may be approached. However, in practice, the efficiency will be lower due to thermal expansion, deviations from a flat density distribution, imperfect kicking fields, and photon-recoil heating during the optical pumping stage. 

To better understand the influence of thermal expansion and the initial density profile, we have developed a one-dimensional analytic model of MOP cooling that is reported in the Appendix. A few key conclusions derived from our model are now summarized. While the MOP cooling sequence progresses, the atomic ensemble's initial standard deviation $\sigma_0$ grows according to $\sigma(t) = \sqrt{\sigma_0^2 + t^2v_{\rm th}^2}$ where $v_{\rm th} = \sqrt{k_{\rm B}T/M}$ is the thermal velocity. Thus, after a MOP cooling sequence, the ensemble has collapsed to a single section, but the sections have grown from their initial size. The total time for one sequence is $t_0+\tau$ and for simplicity let us assume $t_0 \gg 2 \tau$, i.e. the delay time between kicks is much longer than the kicks' combined pulse length. Additionally, a flat distribution of width $2s$ has a standard deviation $\sigma = 2 s / \sqrt{12}$, so the thermally-limited compression factor is approximately $(2J+1) / \sqrt{1 + 12 v_{\rm th}^2 / \Delta v^2}$ where $\Delta v = 2 g_J \mu_{\rm B} G \tau / M \pi$ is the relative kick velocity of adjacent sections and $G$ is the peak field gradient. Thus, the geometric limit $2 J + 1$ is only achieved for $\Delta v \gg \sqrt{12} v_{\rm th}$.

The flat-density initial condition is quite different from that usually encountered with laser cooled atoms in a magneto-optical trap, but turns out to be important for the gains in efficiency that are predicted (details in the Appendix). For example, we predict a compression factor of 1.8 for Li initialized with with a flat density or 1.5 for a Gaussian distribution. For Dy initialized with a flat density, a peak compression factor of 14.4 is observed, while for an initially-Gaussian distribution the factor reduces to 8.6. To obtain a flat ``boxlike” distribution in an experiment, the atoms can first be confined in a magnetic quadrupole trap. An optical box can be created around the atoms using a time-averaged optical dipole potential from beams that are moving rapidly in two dimensions.  Such potentials were created in the past to study optical billiards \cite{raizen2001,friedman2001} and BECs in painted potentials \cite{henderson2009}. After trapping, the box can be adiabatically expanded in three dimensions to a desired size, which will result in a nearly flat density profile of atoms.  The optical setup for state preparation of each segment would enable multiple cycles of cooling in each dimension. Adiabatic expansion would lower the kinetic energy and MOP cooling would compress the cloud spatially.  The process can be dynamically controlled with motorized zoom lenses \cite{kinoshita2005}. 

Ideally, a complete cycle of MOP cooling leaves the velocity distribution of the atomic sample unchanged. This means that the momentum imparted on the atoms in the kick phase must be nulled in the reverse-kick phase. Inhomogeneities in the kicking field will result in a nonzero mean velocity for the cloud. Such inhomogeneities are noticeable in Fig.\,\ref{fig:B_slices}\,({\bf b}) that shows the peak magnetic field gradient in the $x=0$ plane. For instance, atoms near $z=0.25$ will be kicked downward with slightly less force than their upward motion-arresting kick that is applied once they are near $z=0$. Our simulations suggest this effect is negligible for the chosen field and cloud parameters, but it becomes relevant for larger clouds or less uniform field gradients.
Moreover, there exist advanced wiring-design-optimization techniques to generate uniform bias or gradient fields with minimal inductance \cite{wang_improved_2019,turner_minimum_1998}. Though originally developed for magnetic resonance imaging, MOP cooling could benefit from such analyses to improve switching times of the pulsed magnetic fields and increase the uniformity of the required biased gradients, thereby enabling MOP cooling of larger ensembles. 
A possibly more significant mean velocity is incurred due to free fall in the Earth's gravitational field. For example, in the $\sim3.5$\,ms of delay time required for MOP cooling of Cr, the cloud accelerates to nearly 3.5\,cm/s and displaces 0.67 mm. Such velocities and displacements are within typical capture ranges of atomic traps, though some loss upon re-trapping should be expected. To mitigate the free fall effects, one could use the MOP cooling setup to apply an additional uniform kicks against gravity. 

In general, optically pumping the cloud to the same magnetic state is a lossy step due to, for instance, atoms decaying to unobserved trap states. Fortunately, there exist efficient techniques for optical pumping as discussed in \cite{Rochester2016Efficient}, where it is shown that it is possible to perform optical pumping with only one spontaneous photon emitted per atom.
The same physics lets us understand MOP cooling's high phase-space compression efficiency in terms of the photon entropy carried away from the ensemble by spontaneous emission \cite{bartolotta_entropy_2022}.

The entropy associated with the motional degrees of freedom is given by $S_{\rm mo.}=k_{\rm B}\ln (V/V_{\rm Q})$, where $V$ is the phase-space volume and $V_{\rm Q}$ is a reference volume. The ensemble's volume is reduced by a factor of (at most) $2J+1$ at each cooling step, so the maximum entropy reduction per step is $\Delta S_{\rm mo.}=k_{\rm B}\ln{(2J+1)}$. The magnetic kicks are reversible evolution, so they do not produce a net change in entropy. Therefore the entropy change must occur during the optical pumping step. Optical pumping of an unpolarized ensemble to produce a pure state reduces the polarization entropy by $\Delta S_\text{pol.}=k_{\rm B}\ln{(2J+1)}$.
Thus for each step, (1) optical pumping takes unpolarized state to pure state, reducing polarization entropy by $k_{\rm B}\ln{(2J+1)}$; (2) magnetic kicks take the pure state to an unpolarized one by overlapping the sub-ensembles, increasing polarization entropy by $k_{\rm B}\ln{(2J+1)}$ and reducing motional entropy by the same amount.
The cooling must ultimately be limited by recoil heating. 
If single-photon optical pumping efficiency is achieved and spatial compression converted to temperature reduction through adiabatic expansion, the temperature limit will correspond to the recoil temperature $T_{\rm rec}$. For less efficient optical pumping the temperature limit will be higher.

The conceptual significance of MOP cooling is two-fold. First, in species that have already reached quantum degeneracy with the aid of laser cooling, MOP cooling can reduce losses induced by the typical final step of evaporation.
Alternative all-optical approaches have produced degenerate $^{87}{\rm Rb}$ samples while avoiding evaporation: Raman cooling in combination with either optical compression cycles \cite{hu2017creation} or far-off-resonant optical pumping \cite{urvoy2019direct} have reached degeneracy within 300\,ms or 1.4\,s, respectively. The present simulations predict that a single cycle of MOP cooling for Dy will compress the extent of the ensemble along one dimension by a factor of 14.4 within just 3\,ms. This compression increases the number density, and therefore also the phase-space density, by the same factor. Hence, if initially laser cooled to a phase-space density of $\sim10^{-6}$, MOP cooling may bring Dy to degeneracy within six cycles or about 20\,ms. Second, in species lacking a sufficiently-closed two-level cycling transition, MOP cooling provides an alternative technique that only requires optical pumping, coherent population transfer, and pulsed magnetic fields. Here, we have shown that the efficiency of MOP cooling can be increased by leveraging larger internal state spaces. Therefore, a particularly enticing possibility for future work is the application of MOP cooling to ultracold molecules \cite{di2004laser,carr2009cold}. A rich internal-state structure makes laser cooling prohibitively challenging for most, though not all \cite{barau2023blue}, molecular species. On the other hand, molecules may be cooled quite generally to the single-kelvin regime via buffer-gas collisions or supersonic expansion \cite{egorov2002buffer}, and perhaps MOP cooling could push molecular ensembles to yet-higher phase-space densities.

\section{Conclusion}
In this paper, we proposed a highly efficient method for phase-space compression of high-angular-momentum atoms. This work extends an earlier MOP-cooling proposal from Li to general high-angular-momentum atoms. We numerically tested our new MOP-cooling protocol on four atomic species of increasing angular momenta that have each already been cooled using traditional techniques. We find, for example, an impressive compression factor of 14.4 is conceivably attainable in less than 3\,ms for the case of Dy.

\ack
The work of MGR was supported by the Sid W.\,Richardson Foundation. The work of DB was supported by the Deutsche Forschungsgemeinschaft (DFG) - Project ID 423116110 and by the Cluster of Excellence Precision Physics, Fundamental Interactions, and Structure of Matter (PRISMA+ EXC 2118/1) funded by the DFG within the German Excellence Strategy (Project ID 39083149). 

\clearpage
\appendix
\section*{Appendix: One-dimensional model}
\setcounter{section}{1}
In this appendix, we develop a one-dimensional model for MOP cooling. We consider the compression limits imposed by 1) geometry of the initial spatial distribution of atoms and 2) thermal free expansion during the required time-of-flight to complete a MOP cooling cycle. To this end, we begin with the more complicated case of an initially-Gaussian spatial distribution. Ignoring free expansion, we determine an optimal partitioning scheme that splits the initial distribution into $2 J +1$ sections. Then, we consider how the shape of each Gaussian partition changes over the time of flight required to merge the sections. Finally, we repeat the analysis for an initially flat distribution.

\subsection{General setup}
Consider a one-dimensional, classical, non-interacting gas composed of a constant number of atoms with mass $M$ trapped in a potential $V(x)$ and in thermal equilibrium at temperature $T$.  The energy of an atom is $M v^2/2 + V(x)$ when it is at  position $x$ with velocity $v$. The probability for an atom to be found in the phasespace area $(x+dx)\times(v+dv)$ is $\rho(x,v)dxdv$ where the probability density is given by the Boltzmann factor
\begin{equation}
    \rho(x,v) = \rho_0 e^{-[M v^2/2+V(x)]/k_{\rm B}T} ,
\end{equation}
where the constant $\rho_0$ normalizes the density such that it integrates to unity. Therefore, the real-space probability density is given by the marginal
\begin{equation}
    \mathcal{P}(x) =  \int_{-\infty}^{\infty}{\rm d}v  \rho(x,v) 
    =\rho_0 \sqrt{2 \pi v_{\rm th}^2} e ^{-V(x)/k_{\rm B}T} ,
\end{equation}
where we have defined the thermal velocity $v_{\rm th} = \sqrt{k_{\rm B}T/M}$. 

In MOP cooling we imagine slicing the initial density distribution $\mathcal{P}(x)$ into $2 J +1$ sections, where the left-most section is labeled by $m_J=-J$, followed by $-J+1$, and so on to the right most section labeled $m_J=J$.  Then, each section is kicked to a velocity $v_0 \propto -m_J$ such that the distribution is compressed. Our goal is to develop a model allowing us to optimize the MOP cooling protocol for maximum compression. As a figure of merit, we compute the compression factor as the initial-to-final ratio $\mathcal{R}$ of the spatial distribution's standard deviation. We restrict our attention to cases in which $\mathcal{P}$ remains symmetric around $x=0$ so that the distribution's standard deviation is the square-root of its second moment. Therefore, the compression factor is $\mathcal{R} \equiv \sqrt{\langle x^2 \rangle_{\mathcal{P}_{\rm initial}} / \langle x^2 \rangle_{\mathcal{P}_{\rm final}}}$ where
\begin{equation} \label{eq:var_int}
    \langle x^2 \rangle_{\mathcal{P}} = \int_{- \infty}^{\infty} {\rm d}x\, x^2 \mathcal{P}(x) . 
\end{equation}
Note that maximizing $\mathcal{R}$ is equivalent to minimizing $\mathcal{R}^{-2}$, i.e. the ratio of the distribution's final-to-initial variance.

\begin{figure}[t]
\includegraphics[width=\columnwidth]{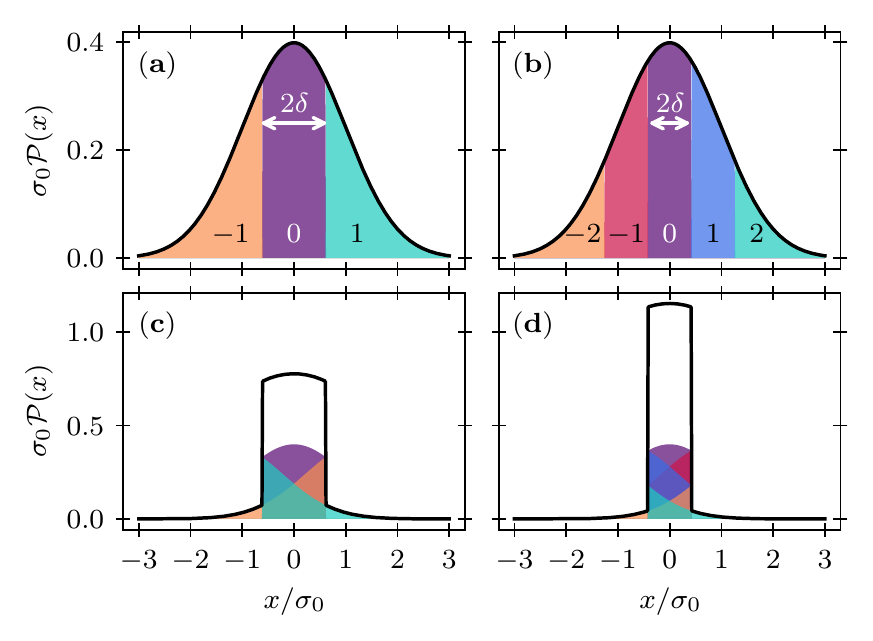}
\caption{The initially Gaussian (a-b) and final (c-d) distributions for MOP cooling of with $J=1$ [(a) and (c)] and $J=2$ [(b) and (d)]. Colors correspond to $m_J$ sections, as labeled in panels (a) and (b), and the black lines mark the total distribution. Here, $\delta$ was chosen according to Equation \eref{eq:gauss_d_model_0T}.} 
 \label{fig:gauss_setup}
\end{figure}

\subsection{Compression limits for a Gaussian distribution}
When trapped in a harmonic potential $V(x)= M \omega_0^2 x^2/2$, the real-space probability density is Gaussian
\begin{equation}
    \mathcal{P}_{\rm G}(x) =   
    \frac{1}{\sqrt{2 \pi \sigma_0^2}} e^{-x^2/2\sigma_0^2} ,
\end{equation}
where we have evaluated the phase-space normalization constant $\rho_0 = 1/2 \pi v_{\rm th} \sigma_0$ and defined the density's variance $\sigma_0^2 = k_{\rm B} T/M \omega_0^2$. The initial standard deviation in this case is simply $\sigma_0=v_{\rm th}/\omega_0$.

Define the the intervals $(a,b)_{m_J}$ that specify the partitioning of the initial distribution as
\begin{equation}
    (a,b)_{m_J} =
    \delta \sigma_0 \cases{
        (-\infty,\, -(2 J + 1)  ] & $m_J = -J$\\
        ((2 m_J -1),\, (2 m_J+1) ] & $|m_J| < J$\\
        ((2J - 1),\, \infty ) & $m_J = J$\\
    } \label{eq:intervals}
\end{equation}
where $2 \delta>0$ represents the distance between adjacent sections in units of $\sigma_0$. At the end of a MOP cooling cycle, these sections have collapsed. Figure \ref{fig:gauss_setup} depicts examples of the initial and final distributions for $J=1,2$. 

\subsubsection{Geometric limit}
If the section collapse occurs much faster than thermal free expansion, then the observed compression is limited by the initial distribution's shape. Typographically, we will use a hat over functions and optimized variables to distinguish this \emph{geometric limit} from the more general results presented later, which account for thermal free expansion. In the present case of an initially-Gaussian distribution, the final distribution is

\begin{equation} \label{eq:P}
    \hat{\mathcal{P}}(x) = 
        \cases{\mathcal{P}_{\rm G}(x-2 J \delta \sigma_0) & $x < -\delta \sigma_0$ \\
        \sum\limits_{m_J=-J}^{J} \mathcal{P}_{\rm G}(x-2 m_J \delta \sigma_0) &$|x| < \delta \sigma_0$\\
        \mathcal{P}_{\rm G}(x+2 J \delta \sigma_0) & $x >\delta \sigma_0$ .}
\end{equation} 

\begin{figure}[t]
\includegraphics[width=\columnwidth]{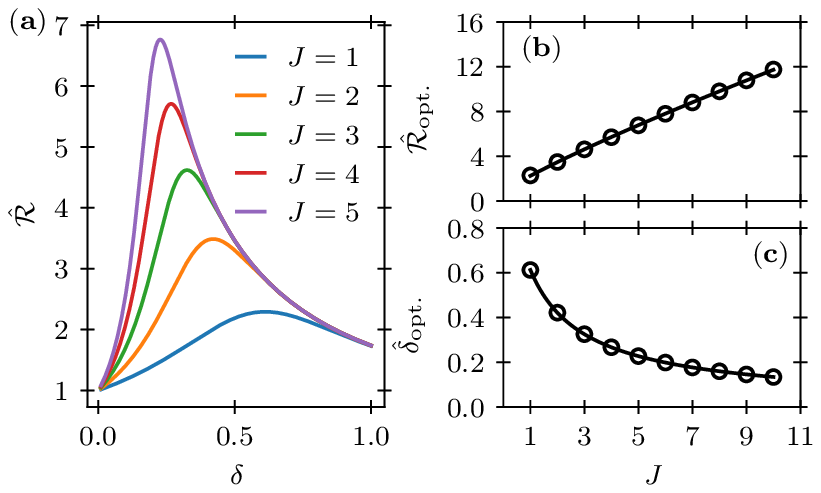}
\caption{(a) The compression factor for various ways of splitting and merging a Gaussian distribution, as given by Equation \eref{eq:gauss_Rd_0T}. We numerically extract the peak compression factor $\hat{\mathcal{R}}_{\rm opt}$ and corresponding relative half-width $\hat{\delta}_{\rm opt.}$ and show their $J$-scaling in panels (b) and (c), respectively. The solid lines in (b) and (c) result from the simultaneous fit to Equations \eref{eq:gauss_R_model_0T} and \eref{eq:gauss_d_model_0T}, respectively.} \label{fig:gauss_mop_opt}
\end{figure}

As for the compression factor $\hat{\mathcal{R}}$, the required integral \eref{eq:var_int} may be evaluated to give
\begin{equation}
    \hat{\mathcal{R}} ^{-2}(\delta, J)= 
    1 - 2 \delta S_1(\delta, J) + 4 \delta^2 \left [ J^2 - S_2(\delta, J) \right ]  ,
    \label{eq:gauss_Rd_0T}
\end{equation}
which is valid for both integer and half-integer $J$ and where we have defined the two parameterized summations
\begin{eqnarray}
    S_1(\delta,J) = \sqrt{\frac{2}{\pi}} \sum_{m_J=-J}^{J-1} \exp \left [ \frac{-\delta^2(2 m_J +1)^2}{2}  \right ] \\
    S_2(\delta,J) = \frac{1}{2}\sum_{m_J=-J}^{J-1} (2 m_J + 1) {\rm Erf} \left [\frac{\delta(2 m_J + 1)}{\sqrt{2}} \right ] .
\end{eqnarray}
In Figure \ref{fig:gauss_mop_opt} (a) we show $\hat{\mathcal{R}}$ as a function of $\delta$ for various $J$. 

Equation \eref{eq:gauss_Rd_0T} provides a way to optimize $\delta$ for a given $J$ by solving $\partial \hat{\mathcal{R}}^{-2}/\partial \delta=0$ for $\delta\equiv \hat{\delta}_{\rm opt}$. Using the identity $S_1^{\prime}(\delta, J)/S_2^{\prime}(\delta, J) = -2 \delta$, where the prime denotes a derivative with respect to $\delta$, we deduce $4 \hat{\delta}_{\rm opt.}[J^2 - S_2(\hat{\delta}_{\rm opt.},J)] - S_1(\hat{\delta}_{\rm opt.},J)=0$. Thus, the corresponding optimal compression factor $\hat{\mathcal{R}}_{\rm opt.}(J) = \hat{\mathcal{R}}(\delta_{\rm opt.},J)$ is given by $\hat{\mathcal{R}}^{-2}_{\rm opt.}=1-\hat{\delta}_{\rm opt.} S_1(\hat{\delta}_{\rm opt.}, J)$. The transcendental form of the above optimization equations precludes a closed form solution. Fortunately, numeric root finding allows us to quickly and accurately obtain $\hat{\delta}_{\rm opt.}$. Empirically, we find that the numerically-optimized $\hat{\delta}_{\rm opt.}(J)$ and the corresponding $\hat{\mathcal{R}}_{\rm opt.}(J)$ are well fit, simultaneously, to the models,
\begin{eqnarray}
    \hat{\mathcal{R}}_{\rm opt.}(J) \approx A_1 (2J + 1)^{B} \label{eq:gauss_R_model_0T} \\
    \hat{\delta}_{\rm opt.}(J) \approx A_2 (2J + A_3)^{-B}\, \label{eq:gauss_d_model_0T},
\end{eqnarray}
as shown in Figure \ref{fig:gauss_mop_opt} (b-c). 
Our fit yields $A_1=0.89$, $A_2=1.82$, $A_3=1.61$, and $B=0.85$. The fitted functions and numeric optimizations agree with an average relative error of 0.33\% for $\hat{\mathcal{R}}_{\rm opt.}$ and 0.08\% for $\hat{\delta}_{\rm opt.}$ for integer $1\leq J \leq10$.

\subsubsection{Thermal limit}
Due to finite magnetic kick strengths, MOP cooling requires a finite time-of-flight to reach peak compression. Due to the ensemble's finite temperature, the spatial and velocity distributions correlate via free expansion, meaning each section becomes more Gaussian-like over the required time-of-flight. Towards a model of MOP cooling including these effects, let us compute the time-dependence of an initially-Gaussian spatial distribution subject to thermal free expansion and translation at a constant kicking velocity $v_0$. We are particularly interested in the time dependence of a subsection of such a distribution that initially occupies the interval $(a,b)$, denoted  $\mathcal{P}_{\rm sec.}\left [x,t; v_0, (a,b) \right ]$. Microscopically, the atom trajectories obey $\dot{v}=0$ and $\dot{x}=v + v_0$, so, by Liouville's theorem, we may write the time-dependent phase-space density as $\rho(x-(v+v_0)t, v)$. The real-space marginal is
%
\begin{eqnarray}
    \mathcal{P}_{\rm sec.}\left [ x,t; v_0, (a,b) \right ] = \frac{1}{2 \pi \sigma_0 v_{\rm th}} \int_{\frac{x-b}{t} -v_0}^{\frac{x-a}{t}-v_0} {\rm d}v\, \\ \nonumber 
     \qquad \times \exp \left ( -\frac{v^2}{ 2 v_{\rm th}^2} - \frac{[x - t (v+v_0)]^2}{2 \sigma_0^2}  \right ) \\ \nonumber
    = \frac{1}{2\sqrt{2 \pi \sigma_0^2 f^2_t}} \exp \left ( \frac{-[x - t v_0 ]^2}{2 \sigma^2_0 f^2_t }\right ) \\ \nonumber
    \qquad  \times \left [ {\rm Erf}\left (\frac{b f_t^2 + t v_0  -x}{\sqrt{2} f_t v_{\rm th} t} \right) -
     {\rm Erf}\left (\frac{a f_t^2 + t v_0  -x}{\sqrt{2} f_t v_{\rm th} t} \right)\right ]
\end{eqnarray}
%
where we have defined $f^2_t = 1 + t^2 v_{\rm th}^2/\sigma_0^2$ and ${\rm Erf}$ is the error function. When $(a,b)=(-\infty,\,\infty)$, then the factor in square brackets becomes $2$, and the result is familiar from time-of-flight analysis of harmonically trapped ensembles. To model MOP cooling, we set the kicking velocity to $v_0 = -m_J \gamma v_{\rm th}$, where $\gamma>0$ measures the kicking strength in units of the thermal velocity, and then sum the distribution sections over the intervals given in Equation \eref{eq:intervals}:
\begin{equation} \label{eq:P_time}
    \mathcal{P}(x, t) = \sum_{m_J =-J}^{J} \mathcal{P}_{\rm sec.}\left [ x,t; -m_J \gamma v_{\rm th}, (a,b)_{m_J} \right ] .
\end{equation}

\begin{figure}[tb]
\includegraphics[width=\columnwidth]{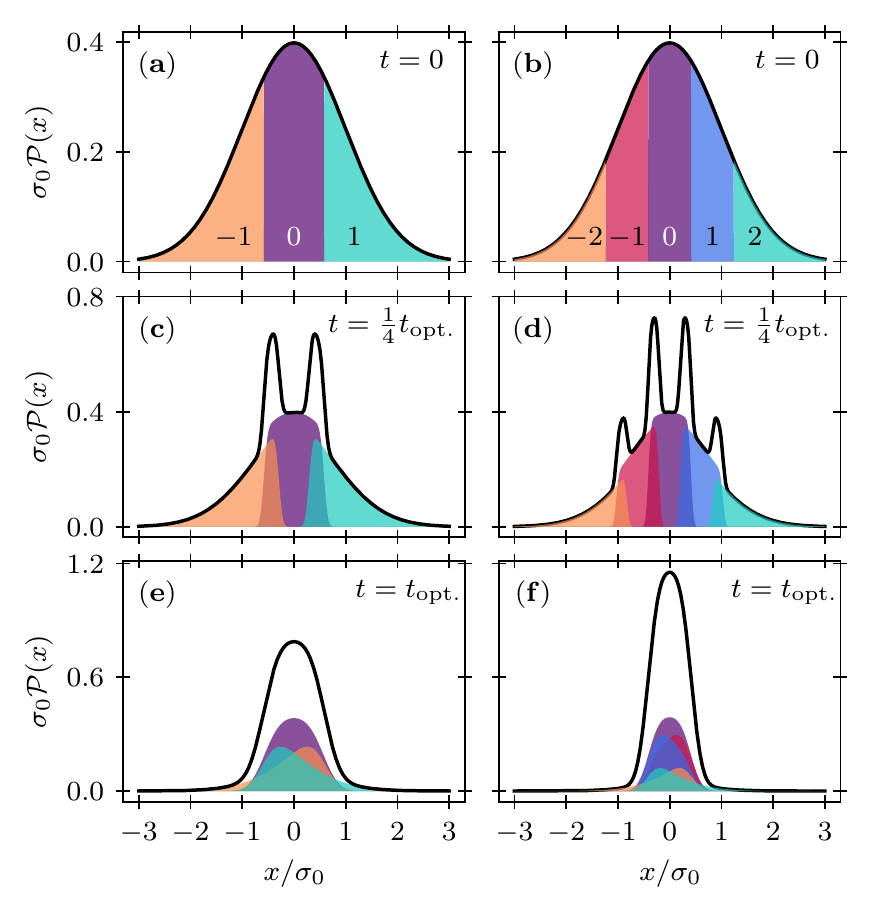}
\caption{Time-dependent MOP cooling  model for a kicking strength of $\gamma=5$ and including effects of free expansion. The initial distributions are shown for (a) $J=1$ and (b) $J=2$. Similarly, panels (c) and (d) show this distributions after a time $t_{\rm opt.}/4$ for $J=1$ and $J=2$, respectively. After a time $t_{\rm opt.}$ the distributions have minimal variance and are shown shown for (e) $J=1$ and (f) $J=2$.} \label{fig:gauss_merge}
\end{figure}

Figure \ref{fig:gauss_merge} shows the time-dependent, real-space probability density for $J=1,2$ and $\gamma=5$ at times $t=0$, $t= t_{\rm opt.}/4$, and $t=t_{\rm opt.}$ where $t_{\rm opt.}\approx 2 \delta \sigma_0 / \gamma v_{\rm th}$ is the optimal delay time after which the sections have collapsed. 

From the probability density \eref{eq:P_time} we may evaluate the compression factor $\mathcal{R}$.  It is convenient to introduce the non-dimensional time $\tau$ such that $t =\tau\sigma_0/v_{\rm th}$. Then, the final result reads
\begin{eqnarray}   \label{eq:gauss_Rtdg}
    \mathcal{R} ^{-2}(\tau, \delta, \gamma, J)= 
    1 &- \tau \gamma S_1(\delta,J) \\ \nonumber &+ \tau^2 \left [ 1  + \gamma^2 J^2 - \gamma^2 S_2(\delta, J) \right ]  .
\end{eqnarray}
The geometric limit \eref{eq:gauss_Rd_0T} is recovered from \eref{eq:gauss_Rtdg} for $\gamma\to \infty$ and $\tau \to 0$ such that $\tau \gamma \to 2 \delta$.

 From \eref{eq:gauss_Rtdg}, the optimal delay time $t_{\rm opt.}=\tau_{\rm opt.} v_{\rm th}/\sigma_0$ may be expressed in closed form upon solving $\partial \mathcal{R}^{-2}/\partial \tau = 0$ for $\tau\equiv \tau_{\rm opt.}$, yielding
\begin{equation}
    \tau_{\rm opt.}(\delta, \gamma, J) =  \frac{\gamma}{2}\frac{S_1(\delta,J)}{1+ \gamma^2 J^2  - \gamma^2S_2(\delta, J)} .
\end{equation}

On the other hand, optimizing $\mathcal{R}$ with respect to the relative half-width $\delta_{\rm opt.}$ for given angular momentum $J$ and kick strength $\gamma$ requires numeric minimization of equation \eref{eq:gauss_Rtdg}. Similarly to the geometric-limit case, analytic optimization implies $4\delta_{\rm opt.}[1 + \gamma^2 J^2 - \gamma^2 S_2(\delta_{\rm opt.},J)] - \gamma^2 S_1(\delta_{\rm opt.},J)=0$ and ${\mathcal{R}}^{-2}_{\rm opt.}(\gamma, J)=1-\delta_{\rm opt.} S_1(\delta_{\rm opt.}, J)$ where $\mathcal{R}_{\rm opt.}(\gamma, J) = \mathcal{R} (\tau_{\rm opt}, \delta_{\rm opt.}, \gamma, J)$.

\subsection{Compression limits for a flat distribution}
When trapped in a box-like potential of width $2s_0$, $V(|x|> s_0) = \infty $ and $V(|x|\leq s_0) = 0$, the ensemble's spatial distribution 
$\mathcal{P}_{\rm F} (|x|<s_0) = 1 / 2s_0$ is flat and has a standard deviation $\sigma_0=s_0/\sqrt{3}$. For a flat distribution, deciding how to partition the $(2J+1)$ sections for MOP cooling is simpler than in the Gaussian case. The flat distribution only has support over the interval $(-s_0,\,s_0)$, so it is intuitive to split the initial distribution into a number $2J+1$ of equally sized sections such that each section contains an equal number of atoms. That is, the intervals of each section are finite and given by 
\begin{equation}
    (a,b)_{m_J} = \left( s_0\frac{ 2 m_J - 1}{2 J +1}, \, s_0\frac{ 2 m_J + 1}{2 J +1} \right ) .
\end{equation}
In the geometric limit, all sections remain flat and merge to a single section of width $2 s_0 / (2 J + 1)$ and hence the geometrically-limited compression factor is $\hat{\mathcal{R}}_{\rm opt.}(J) = 2J+1$.

\begin{figure}[t]
\includegraphics[width=\columnwidth]{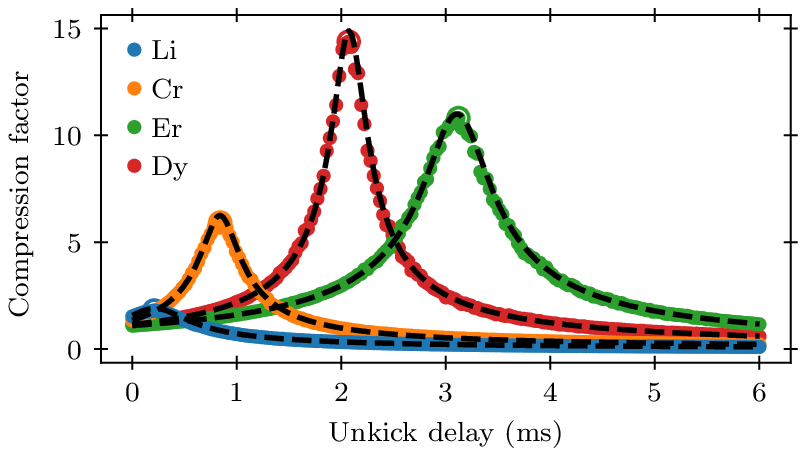}
\caption{The data points are reproduced from Figure \ref{fig:optimize_wait} of the main text and the dashed lines correspond to Equation \eref{eq:flat_Rgt}, to the minus-one-half power, with $\gamma$ and $v_{\rm th}$ set in accordance with the simulations, i.e., no free parameters.}  \label{fig:model_sim_compare}
\end{figure}

 Including the effects of thermal and mean velocity, we find the time-dependent spatial distribution of a section initially occupying the interval $(a,b)$ is 
\begin{eqnarray}
    \mathcal{P}_{\rm sec.}[x,t;v_0,(a,b)] = \frac{1}{4s_0} \bigg [ &{\rm Erf} \left ( \frac{b-x+tv_0}{\sqrt{2} t v_{\rm th}}\right ) \\ \nonumber
    & - {\rm Erf} \left ( \frac{a-x+tv_0}{\sqrt{2} t v_{\rm th}}\right ) \bigg ].
\end{eqnarray}
As in Equation \eref{eq:P_time}, the kicking velocities for each section are $v_0 = -m_J \gamma v_{\rm th}$ and the total distribution is the sum of each section's distribution. The initial variance is given by $\sigma^2_0 = s_0^2/3$. Introducing  the non-dimensional time $\tau=t v_{\rm th} / s_0$ allows us to express the compression factor as
\begin{eqnarray}  \label{eq:flat_Rgt}
     \mathcal{R}^{-2}(\tau, \gamma,J) = 1 &-  \tau \gamma \frac{4 J (J+1)}{(2 J +1)} \\ \nonumber
     &+ \tau^2 \left [ 3 + \gamma^2 J(J+1) \right ]  .
\end{eqnarray}
Equation \eref{eq:flat_Rgt} connects to the simulations discussed in the main text. In Figure \ref{fig:model_sim_compare} we compare Equation \eref{eq:flat_Rgt} to the simulation results presented in Figure \ref{fig:optimize_wait}. We have set $T=25 T_{\rm rec}$ and $\gamma = \Delta v / v_{\rm th}$ where $\Delta v = 2 g_J \mu_{\rm B} G \tau / M \pi$ is the relative kick velocity of adjacent sections and $G$ is the peak field gradient. With no remaining free parameters our one-dimensional model and simulations, which include realistic kicking profiles, are in excellent agreement.

For large enough $\gamma$, we expect the time at which $\mathcal{R}$ is maximized is when all sections have merged at $t_{\rm opt.} \approx 2 s_0 / \gamma v_{\rm th} (2J + 1)$. At this time, the compression factor according to Equation \eref{eq:flat_Rgt} evaluates to
\begin{equation} \label{eq:Rg_flat_approx}
    \mathcal{R}_{\rm opt.}(\gamma, J)  \approx \frac{(2 J + 1) }{\sqrt{1 + 12/\gamma^2}} .
\end{equation}
Equation \eref{eq:Rg_flat_approx} is identical to that given in the discussion section of the main text.
More precisely, solving $\partial \mathcal{R}^{-2} / \partial \tau = 0$ for $\tau\equiv \tau_{\rm opt.}$ gives
\begin{eqnarray}
    \tau_{\rm opt.}(\gamma, J) = \frac{2 \gamma J (J + 1)}{(2 J + 1) \left [ 3 + \gamma^2 J(J+1) \right ]} \label{eq:flat_topt} \\ \nonumber
    = \frac{2}{ \gamma (2 J + 1)} + \mathcal{O}(\gamma^{-2})\\
    \mathcal{R}_{\rm opt.}(\gamma, J) = (2J + 1) \left [\frac{3 + \gamma^2 J(J+1)}{3 +  J(J+1) (\gamma^2 + 12)}\right ]^{1/2} . \label{eq:flat_Ropt}
\end{eqnarray}
The second equality in \eref{eq:flat_topt} comes from a first order expansion in $\gamma^{-1}$ and recovers our $\gamma \to \infty$ expectation. Approximation \eref{eq:Rg_flat_approx} is accurate to better than 10\% (1\%) of Equation \eref{eq:flat_Ropt} when $\gamma\geq2$ and $J=1$ ($J=5$), and accuracy improves monotonically with increasing $\gamma$ and $J$.

The one-dimensional models defined in this appendix are useful to understand the scaling properties of MOP cooling for both Gaussian and flat density profiles. We have outlined techniques for optimization that provide engineering specifications and enable sensitivity analyses for the design of real experimental systems. In the main text, we focused on atomic simulations rather than the models defined here because they not only include thermal free expansion effects, but also three-dimensional, time-dependent kicking forces. In the future, it may be possible to extend our models to include effects like kicking inhomogeneities and finite kicking duration as well as quantum-statistical effects and state preparation details ignored by the main-text simulations.

\section*{References}
\bibliographystyle{unsrt}
\bibliography{bibliography}
\end{document}